\documentstyle{article}

\catcode`@=11
\def\secteqno{\@addtoreset{equation}{section}%
\def\theequation{\thesection.\arabic{equation}}}
\catcode`@=12
\secteqno

\newcommand{\Tb}{\bar{\Theta}}
\newcommand{\C}[1]{{\cal #1}}

\def\CD{{\cal D}}

\def\a{\alpha}              \def\g{\gamma}   \def\d{\delta}

\def\m{\mu}                         \def\p{\pi}
       \def\s{\sigma}             
\def\vp{\varphi}   \def\v{\psi}             

\def\G{\Gamma}             \def\C{\Theta}   
                     
\def\V{\Psi}                \def\T{\Theta}
\newcommand{\nn}{\nonumber}
\newcommand{\ds}{\displaystyle}

\def\tD{{\tilde D}}

\begin{document}
          \hfill 

	  \hfill June, 2002

	  \hfill hep-lat/0206006

          \hfill NIIG-DP-02-5 
\vskip 20mm

\begin{center} 
{\bf \Large
Lattice Chiral Symmetry in Fermionic Interacting Theories and the Antifield Formalism}\\

\vskip 10mm
{\large Yuji\ Igarashi$^a$, Hiroto\ So and Naoya Ukita}\par

\medskip
{\it 
$^a$ Faculty of Education, Niigata University, Niigata 950-2181, Japan\\
Department of Physics, Niigata University, Niigata 950-2181, Japan\\
}

\medskip
\date{\today}
\end{center}
\vskip 10mm

\begin{abstract}
Recently we have discussed realization of an exact chiral symmetry in
theories with self-interacting fermions on the lattice, based upon an
auxiliary field method.  In this paper we describe construction of the
lattice chiral symmetry and discuss its structure in more detail. The
antifield formalism is used to make symmetry consideration more
transparent. We show that the quantum master equation in the antifield
formalism generates all the relevant Ward-Takahashi identities including
a Ginsparg-Wilson relation for interacting theories. Solutions of the
quantum master equation are obtained in a closed form, but the resulting
actions are found to be singular. Canonical transformations are used to
obtain four types of regular actions. Two of them may define consistent
quantum theories. Their Yukawa couplings are the same as those obtained
by using the chiral decomposition in the free field algebra.  Inclusion
of the complete set of the auxiliary fields is briefly discussed.

\end{abstract}




\setcounter{footnote}{0}

\newpage

\section{Introduction}
Recently much attention has been paid to new realization of symmetries
which are not compatible, in ordinary sense, with regularizations.  The
discovery of the exact chiral symmetry on the lattice
\cite{gw}-\cite{her} may be regarded as the prototype of such
realization. (See, for example, ref.\cite{reviews} for reviews of recent
development.) The need for considering regularization-dependent
symmetries is not restricted to lattice theories: A closely related
issue in continuum theories is to realize gauge symmetries in the
Wilsonian renormalization group (RG) \cite{wk}, which introduces an
infrared cutoff to define the Wilsonian effective action.

In the new realization of a symmetry, such as the lattice chiral
symmetry or the gauge symmetry along the RG flow, symmetry
transformation inevitably depends on cutoff parameter so as to make
reconciliation with a regularization. For such a theory, neither an
action nor path integral measure may remain invariant under the
transformation. One has to determine the action and the symmetry
transformation at the same time in such a way that total change arising
from the action and the measure vanish.  It is therefore difficult in
general to give a nonperturbative formulation of the symmetry in the
presence of interactions.

For the chiral symmetry on the lattice, the gauge interactions were
incorporated into the Dirac operator in vector gauge theories. Recently,
 chiral gauge theories with anomaly-free fermion multiplets have been
constructed in the abelian case \cite{lu2}. In these theories, 
the Ginsparg-Wilson (GW) relation \cite{gw}, the crucial algebraic identity 
in formulation of the
symmetry, takes the same form as that in free field theories. The
significance and consequences of this identity has been extensively
discussed \cite{neu1}-\cite{reviews}\cite{lu2}-\cite{fi}. However, it
seems to be not clear how the GW relation should be generalized for
describing other interactions.

In a previous work \cite{isu}, referred to as I hereafter, we discussed
a lattice chiral symmetry in theories with fermionic self-interactions,
introducing a scalar and a pseudoscalar fields as auxiliary fields. A GW
relation for these theories was given. It depends on the Yukawa
couplings to the auxiliary fields. The Yukawa interactions we obtained
take different form from those considered in the literature 
\cite{reviews}\cite{Naya}\cite{chand}\cite{in}\cite{fi}.  The
chiral transformation obtained from the GW relation becomes
nonlinear. Under the transformation, none of the kinetic term of the
Dirac fields, the Yukawa coupling term and the functional measure
remains invariant. Although we gave chiral invariant partition function,
these peculiar properties of the chiral symmetry remained to be studied.

The purpose of this paper is twofold: First, we consider construction
of the lattice chiral symmetry discussed in I in more detail, using a
framework of the antifield formalism originally developed by Batalin
and Vilkovisky \cite{bv}. Second, besides the peculiar properties of
the chiral symmetry described above, we argue that the the actions
obtained in I are singular. In order to have regular actions, we introduce 
new dynamical variables, and it enables us to reconstruct the
chiral symmetry. 

The reason why we use the antifield formalism to describe the lattice
chiral symmetry is as follows: The formalism has been recognized as the
most general and powerful method of BRS quantization for theories with
local as well as global symmetries. For regularization-dependent
symmetries, it is a nontrivial problem to derive the Ward-Takahashi (WT)
identities, and is certainly desirable to develop a general method for
doing this task both in lattice and continuum theories.  We believe that
the antifield formalism is a strong candidate for the method. Actually,
it has been applied to regularization-dependent symmetries realized
along the RG flow \cite{iis1}\cite{iis2}. In the antifield formalism,
the presence of exact symmetries, irrespective of whether they are
regularization-dependent or not, is expressed by the quantum master
equation (QME) \cite{bv}, $\Sigma =0$.

For the fermionic theories we consider, antifields are introduced at the
stage of the block transformation \cite{gw} from a microscopic
theory on a fine lattice to its macroscopic counterpart on a coarse
lattice. In performing the transformation, we also use the auxiliary
field method discussed in I.  A scalar and a pseudo-scalar fields are
introduced again to make an effective description of the fermionic
self-interactions. Introduction of antifields for the Dirac and the
auxiliary fields makes it possible to encode the chiral transformation
in a natural way. This should be compared with the somehow ad hoc way of
postulating the transformation rule in the conventional approach.

Since the fermionic interaction terms are replaced by the Yukawa term
and a potential of the auxiliary fields, the fermionic sector in the
macroscopic action is linearized. The price for the use of the auxiliary
field method is that we only obtain a more weaker condition,
vanishing of the expectation value $<\Sigma> =0$, rather than the QME 
$\Sigma =0$. 
Although the condition $<\Sigma> =0$
allows a wide class of solutions, we consider here only the solutions to
the QME.  We show that the QME generates all the relevant WT identities
including the GW relation obtained in I.  Under suitable assumptions,
the QME can be solved in a closed form.  The solutions are used to
construct four types of chiral invariant partition functions.

The actions constructed with these sets of solutions turn out to have
singularities in the Yukawa couplings as well as in the auxiliary field
potential. The singularities\footnote{A similar kind
of singularities has been discussed in different context in \cite{fi}.} 
in the Yukawa couplings arise at the
momentum regions where would-be species doublers appear.  
In order to remove the singularities, we perform canonical
transformations in the space of fields and antifields. Among four sets
of the transformed regular actions, two of them contain massless doubler
modes which decouple to the auxiliary fields. This decoupling occurs at
tree level but could not be stable due to the quantum corrections. In
other two sets of the actions, the doubler modes become massive, and
decoupling to the auxiliary fields is ensured by the chiral symmetry. In
these actions, the kinetic term of the Dirac fields, the Yukawa coupling
term and the functional measure are all chiral invariant. The chiral
transformation of the new variables takes the same form as the one for
the free field theory. The Yukawa couplings coincide with those
discussed in \cite{reviews}\cite{chand}\cite{in}\cite{fi}. They can be
obtained by using the chiral decomposition in the free field algebra
\cite{reviews}\cite{Naya}\cite{iis2}.  The actions obey the classical
master equation rather than the QME.

As for the auxiliary fields, we restrict ourselves mostly to a scalar
and a pseudoscalar fields. We may include other auxiliary fields in a
similar manner. We try here to give a formal argument of inclusion of
the complete set of the fields with multi-flavor, though the constructed
actions suffer from the singularities discussed above.

This paper is organized as follows.  In section 2, we describe
construction of macroscopic action for fermionic system introducing
scalar and pseudoscalar auxiliary fields. Then, the block transformation
\cite{gw} is reconsidered in the antifield formalism. In section 3, the
QME is derived. Under suitable assumptions, we show that it yields the
WT identities discussed in I.  Two sets of particular solutions of the
QME are given. Using these non-perturbative solutions, we give chiral
invariant partition functions on the coarse lattice in section 4. In
section 5, reconstruction of the chiral symmetry using canonical
transformations is discussed. We perform inclusion of the complete set
of the auxiliary fields in section 6. The section 7 is devoted to
summary and discussion. Derivation of some formulae is given in
Appendix.

\section{Macroscopic action in the antifield formalism}
In this section, we first construct a macroscopic action without
introducing the antifields. Although this was discussed in I, a brief
summary of the construction is given to make the present work
self-contained. We then reconsider our microscopic as well as macroscopic
theories in the antifield formalism, and give a phase-space extension of
the block transformation. 
\subsection{Construction of macroscopic action}
Let $A_{\rm c}[\v_{\rm c},\bar{\v}_{\rm c}]$ be a microscopic action of
the Dirac fields $\v_{\rm c}(x),\bar{\v}_{\rm c}(x)$. The fields are
defined on a $d$ (even) dimensional fine lattice whose positions are
labeled by $x$. For simplicity, they are assumed to carry a single
flavor.  The microscopic action describes a certain class of fermionic
self-interactions. Let $A[\V,\bar{\V}]$ be an effective action of the
Dirac fields $\V_{n},\bar{\V}_{n}$ defined on a coarse lattice.  Indices
$n, m$ are used for labeling sites of the lattice. The macroscopic
action is obtained from the microscopic action via the block transformation
\begin{eqnarray}
 e^{-A[\V,\bar{\V}]} & \equiv &
 \int{\cal D}\v_{\rm c}{\cal D}\bar{\v}_{\rm c}\ 
 e^{- A_{\rm c}[\v_{\rm c},\bar{\v}_{\rm c}]
    - \sum_{n}(\bar{\V}_{n}-\bar{B}_{n})\a(\V_{n}-B_{n})},
 \label{2.3}
\end{eqnarray}
where $\a$ is a constant parameter proportional to inverse of the coarse
lattice spacing $a$, $\a \propto a^{-1}$. The gaussian integral in
(\ref{2.3}) relates the macroscopic fields $\V_{n},\bar{\V}_{n}$ to the
block variables defined by
\begin{eqnarray}
 \left\{
 \begin{array}{ccl}
   B_n & \equiv &
   \int d^{d}x\ f_n(x) \v_{\rm c}(x) \\
   \bar{B}_n & \equiv &
   \int d^{d}x\ \bar{\v}_{\rm c}(x) f_{n}^{\ast}(x)
 \end{array}
 \right. ,
 \label{2.4}
\end{eqnarray}
where $f_{n}(x)$ is an appropriate function for coarse graining. It is
normalized as $\int d^{d}x\ f_{n}^{\ast}(x) f_{m}(x) = \d_{nm}$.

The path-integral over the microscopic fields in (\ref{2.3}) will
generate fermionic self-interaction terms in the macroscopic action
$A[\V,\bar{\V}]$. Instead of dealing with such terms directly, we
introduce some auxiliary fields on the coarse lattice to describe the
fermionic interactions. The maximal number of the auxiliary fields to be
introduced is equal to the dimension of the Clifford algebra, i.e.,
$2^d$ for the $d$ (even)-dimensional Dirac fields.  In section 6, we
discuss the inclusion of the complete set of the auxiliary fields. For
simplicity, we restrict ourselves here to a scalar $\s_n$ and a
pseudoscalar field $\p_n$, because the scalar and pseudoscalar
interactions are recognized as the most important couplings to describe
chiral symmetry and its spontaneous breaking in the effective theory.
The macroscopic action we consider then takes of the form
\begin{eqnarray}
 A[\V,\bar{\V}] &=& \sum_{nm}\biggl\{\bar{\V}_{n} (D_{0})_{nm} \V_{m} \nonumber
\\
&{}& + 
 V[\bar{\V}_{n} (\d_{nm} + h(\nabla)_{nm})\V_{m}, \bar{\V}_{n} \gamma_{5}
 (\d_{nm} + h(\nabla)_{nm})\V_{m}]\biggr\} ,
\label{2.5}
\end{eqnarray}
where $D_{0}$ is the Dirac operator for the kinetic term, and $ V$
denotes fermionic interactions which consist of contact terms as well as
non-contact ones with the difference operators $h(\nabla)_{nm}$.  We may
obtain the action (\ref{2.5}) by performing integration over the
auxiliary fields in a new macroscopic action:
\begin{eqnarray}
 e^{ -A[\V,\bar{\V}]}  &\equiv&
 \int{\cal D}\p{\cal D}\s  \nonumber\\ 
&{}&\times ~ 
 e^{- \sum_{nm}\bar{\V}_{n}\left(D_{0} + \d + h(\nabla)\right)_{nm}
 (i\g_5\p + \s )_{m}\V_{m}- A_{X}[\p,\s]}.
 \label{2.6}
\end{eqnarray}
It is noted that the Dirac fields appear only bilinearly in the new action. 
All the fermionic interactions are cast into the Yukawa couplings
 with the auxiliary fields and the potential term $A_{X}[\p,\s]$.
In summary, the block transformation is given by 
\begin{eqnarray}
 \lefteqn{\int{\cal D}\p{\cal D}\s \  
  e^{- \sum_{nm}\bar{\V}_{n}{\tilde D}(\p,\s)_{nm} \V_{m}
     - A_{X}[\p,\s]}} & & \nonumber \\ 
 & \hspace{2cm} = & 
 \int{\cal D}\v_{\rm c}{\cal D}\bar{\v}_{\rm c}\  
  e^{- A_{\rm c}[\v_{\rm c},\bar{\v}_{\rm c}]
     - \sum_{n}(\bar{\V}_{n}-\bar{B}_{n})\a(\V_{n}-B_{n})} ,
 \label{2.7}
\end{eqnarray}
with the total Dirac operator
\begin{eqnarray}
{\tilde D}(\p,\s)_{nm} &=& (D_{0})_{nm} + \left(\d + h(\nabla)\right)_{nm}
(i\g_5\p + \s)_{m} \nonumber\\
&\equiv& D(\p,\s)_{nm} + (i\g_5\p + \s)_{n}.    
\label{2.8}
\end{eqnarray}
Here ${\tilde D}(\p,\s)$ is assumed to be at most linear in $\pi$
and $\s$. We now reconsider the block transformation (\ref{2.7}) in 
the antifield formalism.

\subsection{The antifield formalism and the block transformation}
The antifield formalism \cite{bv} describes any local or global symmetry
as a ``BRS'' symmetry. It defines a kind of ``canonical structure'' for a
given action of fields by adding their ``momentum variables'' called
antifields. In our previous papers \cite{iis1}\cite{iis2}, the formalism
has been used for realization of symmetries along the Wilsonian RG flow.
The purpose of this subsection is to give a lattice version of the
formalism in the context of chiral symmetry.

In the antifield formalism, the chiral transformation in the microscopic
theory takes the form of BRS transformation:
\begin{eqnarray}
\d_{B} \v_{\rm c}(x) &=& i~C~\g_{5} \v_{\rm c}(x), \nonumber \\
\d_{B} \bar{\v}_{\rm c}(x) &=& i~C~\bar{\v}_{\rm c}(x) \g_{5}, 
\label{3.1}
\end{eqnarray}
where $C$ is a constant ghost. It is Grassmann odd, therefore $C^{2}=0$. 
For the Dirac fields $\phi^{a}\equiv \{\v_{\rm c}, \bar{\v}_{\rm c}\}$,
one introduces anti-Dirac fields $\phi^{*}_{a} \equiv \{\v_{\rm c}^{*},
\bar{\v}_{\rm c}^{*}\}$.  Although the antifields are unphysical, they
play an important r\^{o}le for encoding chiral symmetry, and should be
eliminated only at the final stage of our calculation. In order to
include the antifields, one considers an extended
microscopic action,
\begin{eqnarray}
S_{\rm c}[\phi, \phi^{*}] \equiv A_{\rm c}[\v_{\rm c},\bar{\v}_{\rm c}] + 
\int d^{d} x [\v_{\rm c}^{*}(x) \d_{B} \v_{\rm c}(x) + 
\d_{B} \bar{\v}_{\rm c}(x) \bar{\v}_{\rm c}^{*}(x)].
\label{3.2}
\end{eqnarray}
It is noted here that the BRS transformation operator $\d_{B}$ is Grassmann
odd and carries one unit of ghost number. Therefore, the antifields
$\phi^{*}_{a}$ regarded as source terms for $\d_{B} \phi^{a}$ are Grassmann
even and carry ghost number $-1$. The canonical structure in the theory
with antifields is specified by the antibracket. For any functions
$F[\phi, \phi^{*}]$ and $G[\phi, \phi^{*}]$, it is defined by
\begin{eqnarray}
\left(F,~G\right)_{\phi}
&=& \int d^{d} x\biggl[
\frac{{\partial}^{r} F}{\partial \psi_{\rm c}(x)} 
\frac{{\partial}^{l} G}{\partial \psi_{\rm c}^{*}(x)}
-\frac{{\partial}^{r} F}{\partial \psi_{\rm c}^{*}(x)} 
\frac{{\partial}^{l} G}{\partial \psi_{\rm c}(x)}\nonumber\\
&~~~&+ \frac{{\partial}^{r} F}{\partial \bar{\psi}_{\rm c}(x)} 
\frac{{\partial}^{l} G}{\partial \bar{\psi}_{\rm c}^{*}(x)}
-\frac{{\partial}^{r} F}{\partial \bar{\psi}_{\rm c}^{*}(x)} 
\frac{{\partial}^{l} G}{\partial \bar{\psi}_{\rm c}(x)}\biggr].
\label{3.3}
\end{eqnarray}
The chiral transformation of $F$ is described as
\begin{eqnarray}
\d_{B} F = (F, S_{c})_{\phi}.
\label{3.4}
\end{eqnarray}
Note that this is an operation from the right.  If the
original action $A_{\rm c}[\v_{\rm c},\bar{\v}_{\rm c}]$ is chiral
invariant $\d_{B} A_{\rm c}=0$, the extended action $S_{\rm c}[\phi,
\phi^{*}]$ is so, too. It is expressed by the classical master equation, 
\begin{eqnarray}
(S_{c},~ S_{c})_{\phi} =0.
\label{3.5}
\end{eqnarray}

For the Dirac fields on the coarse lattice, we introduce their
anti-fields $\V^{*}_{n}$ and $ \bar\V^{*}_{n}$. We also include  
anti-auxiliary fields: $\pi^{*}_{n}$ and ${\s}^{*}_{n}$.  Let $\Phi^{A}
\equiv \{\V_{n}, \bar{\V}_{n}, \pi_{n}, \s_{n}\}$ be all the fields on
the coarse lattice, and $\Phi^{*}_{A} \equiv \{\V^{*}_{n},
\bar{\V}^{*}_{n}, \pi^{*}_{n}, {\s}^{*}_{n}\}$ be their antifields.
Then, a phase-space extension of (\ref{2.7}) is given by
\begin{eqnarray}
\lefteqn{
\int{\cal D}\pi {\cal D} \s{\cal D}\pi^{*} {\cal D}\s^{*}
 \prod_{n}\d (\pi_n^{*})\d (\s_n^{*})~
 e^{-S[\Phi, \Phi^{*}]}} \nn \\ 
 &=& 
\int{\cal D}\phi {\cal D}\phi^{*}
\prod_{x}\d 
\biggl(\sum_{n}\V_{n}^{*}f_{n}(x)-\v_{c}^{*}(x)\biggr)
\d 
\biggl(\sum_{n}\bar{\V}_{n}^{*}f_{n}^{*}(x)-{\bar\v}_{c}^{*}(x)\biggr)
\nn \\&{}&\times
 e^{- S_{\rm c}^{\rm total}[\phi, \phi^{*}]},
\label{3.6}
\end{eqnarray}
where the block transformation for the antifield sector is
described  
by using the $\delta$ functions. 
The total microscopic action in (\ref{3.6}),
\begin{eqnarray}
S_{\rm c}^{\rm total}[\phi, \phi^{*}] &\equiv& 
S_{\rm c}[\phi, \phi^{*}]
    + \sum_{n}(\bar{\V}_{n}-\bar{B}_{n})\a(\V_{n}-B_{n}),
\label{3.7}
\end{eqnarray}
has terms linear in the microscopic Dirac fields:
\begin{eqnarray}
\int d^{d}x \sum_{n}\bigl(f_{n}^{*}(x)\bar{\v}_{\rm c}(x) \a 
({\V} + i~C\g_{5} {\a}^{-1} \bar\V^{*})_{n}  
+ f_{n}(x) (\bar{\V}- \V^{*}i~C \g_{5}{\a}^{-1})_{n}
  \a {\v}_{\rm c}(x) \bigr).
\nonumber
\end{eqnarray}
It implies that the effective source terms 
for $\bar{\v}_{\rm c}$ and ${\v}_{\rm c}$ are proportional to 
$({\V} + i~C\g_{5} {\a}^{-1} \bar\V^{*})_{n}$ and 
$(\bar{\V}- \V^{*}i~C \g_{5}{\a}^{-1})_{n}$, respectively. 
Thus, we find that the macroscopic action takes of the form
\begin{eqnarray}
S[\Phi, \Phi^{*}]& =& \sum_{nm}
(\bar{\V}- \V^{*}i~C \g_{5}{\a}^{-1})_{n}({\tilde D}
(\pi,\s) - \a)_{nm} ({\V} + i~C\g_{5} {\a}^{-1} \bar\V^{*})_{m}\nonumber\\
&{}& + \sum_{n} \bar{\V}_{n} \a {\V}_{n} + A_{X}[\pi,\s] + 
 \sum_{n}(\pi^{*}_{n} \d_{B} \pi_{n} + \s^{*}_{n} \d_{B} \s_{n}),
\label{3.8}
\end{eqnarray}
where the anti-auxiliary fields are included. They are multiplied by the
BRS transformed auxiliary fields, $\d_{B} \pi_{n}$ and
$\d_{B} \s_{n}$, which are to be determined later.  We have used the
total Dirac operator ${\tilde D}(\pi,\s)$ given in (\ref{2.8}).

It is noted that the chiral transformation for the Dirac fields
$\V,~\bar{\V}$ is automatically encoded due to the presence of the
anti-Dirac fields $\V^{*},~\bar\V^{*}$ in (\ref{3.8}):
\begin{eqnarray}
\d_{B} {\V}_{n} &=& \left({\V}_{n},~S[\Phi, \Phi^{*}]\right)_{\Phi}
= i~C\g_{5}\left(1 - {\a}^{-1}{\tilde D}\right)_{nm}{\V}_{m},\nonumber\\
\d_{B} \bar{\V}_{n} &=& \left(\bar{\V}_{n},~S[\Phi, \Phi^{*}]\right)_{\Phi} 
= i~C \bar{\V}_{m}\left(1 - {\a}^{-1}{\tilde D}\right)_{mn}\g_{5},
\label{3.9}
\end{eqnarray}
where $(~,~)_{\Phi}$ denotes the antibracket for the macroscopic sector. It
is given by
\begin{eqnarray}
\left(F,~G\right)_{\Phi}
&=& \left(F,~G\right)_{D} + \left(F,~G\right)_{X},\nonumber\\
\left(F,~G\right)_{D} &=& \sum_{n}\biggl[
\frac{{\partial}^{r} F}{\partial \Psi_{n}} 
\frac{{\partial}^{l} G}{\partial \Psi_{n}^{*}}
-\frac{{\partial}^{r} F}{\partial \Psi_{n}^{*}} 
\frac{{\partial}^{l} G}{\partial \Psi_{n}}\nonumber\\
&~~~&+ \frac{{\partial}^{r} F}{\partial \bar{\Psi}_{n}} 
\frac{{\partial}^{l} G}{\partial \bar{\Psi}_{n}^{*}}
-\frac{{\partial}^{r} F}{\partial \bar{\Psi}_{n}^{*}} 
\frac{{\partial}^{l} G}{\partial \bar{\Psi}_{n}}\biggr],\nonumber\\
\left(F,~G\right)_{X} &=& \sum_{n}\biggl[
\frac{{\partial}^{r} F}{\partial {\p}_{n}} 
\frac{{\partial}^{l} G}{\partial \p_{n}^{*}}
- \frac{{\partial}^{r} F}{\partial \p_{n}^{*}} 
\frac{{\partial}^{l} G}{\partial \p_{n}}\nonumber\\
&~~~&+\frac{{\partial}^{r} F}{\partial {\s}_{n}} 
\frac{{\partial}^{l} G}{\partial \s_{n}^{*}}
- \frac{{\partial}^{r} F}{\partial \s_{n}^{*}} 
\frac{{\partial}^{l} G}{\partial \s_{n}}
\biggr].
\label{3.10}
\end{eqnarray}

In the antifield formalism, one can use canonical transformations for
the phase-space variables. They are defined as transformations from
$\{\Phi^{A},~\Phi^{*}_{A}\}$ to $\{\Phi^{\prime A},~\Phi^{*\prime }_{A}\}$
that render the antibrackets invariant: $(F,~G)_{\Phi}=(F,~G)_{\Phi'}$.
One thing which should be remarked is that the path-integral measure
${\cal D}\Phi {\cal D}\Phi^{*}$ is not left invariant in general under
the canonical transformations. Since there is no Liouville measure in
the phase-space, one has to take account of the associated Jacobian
factor in quantum theory.  

We can rewrite the fermionic part of the action
in (\ref{3.8}) by performing a canonical transformation,
\begin{eqnarray}
\V_{n}^{\prime} &=&\V_{n} + i~ C \g_{5} \a^{-1} \bar{\V}_{n}^{*},\nonumber\\
 \bar{\V}_{n}^{\prime} &=&  \bar{\V}_{n} +\V_{n}^{*} i~C  \g_{5} \a^{-1},\nonumber\\
\V_{n}^{*\prime} &=& \V_{n}^{*}, \nonumber\\
\bar{\V}_{n}^{*\prime} &=& \bar{\V}_{n}^{*}.  
\label{3.17}
\end{eqnarray}
It is easy to see that the Jacobian factor associated with this
canonical transformation is trivial. Hereafter, we use the new set of
variables in construction of the lattice chiral symmetry, and represent
it by $\{ \V, \bar{\V}, \V^{*}, \bar{\V}^{*}\}$ removing primes.  Then,
using these variables, the macroscopic
extended action (\ref{3.8}) is expressed as
\begin{eqnarray}
S[\Phi,~\Phi^{*}]
&=& S_{D} + S_{X},\nonumber\\
S_{D}&=& \sum_{nm}\bar{\V}_{n}{\tilde D}(\pi,\s)_{nm}\V_{m} \nn \\ &&
+ 
\sum_{nm}\biggl[
\V_{n}^{*}i~C \g_{5}\bigl(1 -2{\a}^{-1} {\tilde D}\bigr)_{nm}
\V_{m} -  \bar{\V}_{n} i~C \g_{5} \d_{nm}\bar{\V}_{m}^{*} 
\biggr],\nonumber\\
S_{X} &=& A_{X}[\pi,\s] + 
 \sum_{n}(\pi^{*}_{n} \d_{B} \pi_{n} + \s^{*}_{n} \d_{B} \s_{n}).
\label{3.18}
\end{eqnarray}
It leads to the asymmetric form of the chiral transformation:
\begin{eqnarray}
\d_{B} \V_{n} &=& i~C \g_{5}\left(1 -2 {\a}^{-1}{\tilde D}\right)_{nm} 
\V_{m},\nonumber\\ 
\d_{B} \bar{\V}_{n} &=& i~C  \bar{\V}_{n} \g_{5}.
\label{3.19}
\end{eqnarray}
Here $\bar{\V}$ obeys the standard chiral transformation, while $\V$
does not. Instead, we may consider the chiral transformation
\begin{eqnarray}
\d_{B} \V_{n} &=& i~C \g_{5}\V_{n},\nonumber\\ 
\d_{B} \bar{\V}_{n} &=& i~C  \bar{\V}_{m} \left(1 -2 {\a}^{-1}{\tilde D}\right)_{mn}\g_{5},
\label{3.191}
\end{eqnarray}
where  $\V$ obeys the standard chiral transformation, while $\bar{\V}$
does not. The Dirac action which leads to (\ref{3.191}) is given by
\begin{eqnarray}
S_{D}&=& \sum_{nm}\bar{\V}_{n}{\tilde D}(\pi,\s)_{nm}\V_{m} \nn \\ &&
+ 
\sum_{nm}\biggl[
\V_{n}^{*}i~C \g_{5}\d_{nm}
\V_{m} -  \bar{\V}_{n} i~C \left(1 -2 {\a}^{-1}{\tilde D}\right)_{nm} \g_{5} 
\bar{\V}_{m}^{*} 
\biggr],
\label{3.181}
\end{eqnarray}
which can be obtained from (\ref{3.8}) via another canonical transformation.

In this section, we have given the block transformation (\ref{3.6})
in the antifield formalism. It relates symmetry properties in the
microscopic theory to those in the macroscopic theory. After a canonical
transformation, the extended action (\ref{3.18}) or that with the action 
for the Dirac fields (\ref{3.181}) has been obtained. We
consider below the WT identities for this action. 

\section{The QME and its solutions}
In the antifield formalism, the basic object which detects the presence
of symmetry in a given quantum system is the WT operator. For a
path-integral $\int {\cal D} \vp e^{-W[\vp, \vp^{*}]}$ with an action
$W[\vp, \vp^{*}]$, the WT operator is defined as $\Sigma[\vp, \vp^{*}] =
(W,~W)_{\vp}/2 - \Delta_{\vp}W = e^{W} \Delta_{\vp} e^{-W}$, where
$\Delta$ denotes the ``divergence'' operator whose explicit expression
is given below. The WT operator $\Sigma[\vp, \vp^{*}]$ can be
interpreted as follows: Consider a change of variables $\vp \to \vp +
(\vp, W)_{\vp}$. It induces changes in the action by $(W,W)_{\vp}/2$
and those arising from the functional measure by
$\Delta_{\vp}W$. Invariance of the path-integral requires cancellation
of these two contributions: $\Sigma[\vp, \vp^{*}] = 0$. This is the QME,
which ensures the presence of BRS symmetry in the quantum system. In
this section, we derive the QME in our macroscopic theory, and then
solve it.

\subsection{The QME in the macroscopic theory}
For the microscopic action, the WT operator reads
\begin{eqnarray}
\Sigma [\phi, \phi^{*}] \equiv \frac{1}{2}(S_{c}, S_{c})_{\phi} 
- \Delta_{\phi} S_{c}, 
\label{4.1}
\end{eqnarray}
where the $\Delta$-derivative is given by
\begin{eqnarray}
\Delta_{\phi}= \int d^{d}x \biggl[\frac{{\partial}^{r} }
{\partial \psi_{c}(x)} 
\frac{{\partial}^{r} }{\partial \psi^{*}_{c}(x)}
-\frac{{\partial}^{r} }
{\partial {\bar\psi}_{c}(x)} 
\frac{{\partial}^{r} }{\partial {\bar\psi}^{*}_{c}(x)}
\biggr].
\label{4.2}
\end{eqnarray}

We now discuss the relation between the WT operator for the microscopic 
action and that for the macroscopic action. To this end, we consider 
the functional average of the WT operator in the microscopic theory 
\cite{iis1} 
\begin{eqnarray}
\lefteqn{ \left<\Sigma[\phi,\phi^{*}]\right>_{\phi} } \nn \\  
&=& \biggl[\int{\cal D}\phi {\cal D}\phi^{*}\prod_{x}\d 
\bigl(\sum_{n}\V_{n}^{*}f_{n}(x)-\v_{c}^{*}(x)\bigr)
\d 
\bigl(\sum_{n}\bar{\V}_{n}^{*}f_{n}^{*}(x)-{\bar\v}_{c}^{*}(x)\bigr)
\nonumber\\
&{}& \times ~e^{- (S_{\rm c}^{\rm total}- S_{c})}\Delta_{\phi}e^{ - S_{c}}
\biggr]
\biggl[\int{\cal D}\phi {\cal D}\phi^{*}\prod_{x}\d 
\bigl(\sum_{n}\V_{n}^{*}f_{n}(x)-\v_{c}^{*}(x)\bigr)\nonumber\\
&{}& \times \d 
\bigl(\sum_{n}\bar{\V}_{n}^{*}f_{n}^{*}(x)-{\bar\v}_{c}^{*}(x)\bigr)
e^{- S_{\rm c}^{\rm total}}\biggr]^{-1}, 
\label{4.3}
\end{eqnarray}
where the actions $ S_{c}$ and $S_{\rm c}^{\rm total}$ are those defined 
in (\ref{3.2}) and (\ref{3.7}). Performing integration by parts in
(\ref{4.3}) and using (\ref{3.6}), one obtains
\begin{eqnarray}
\left<\Sigma [\phi, \phi^{*}]\right>_{\phi} 
&=&\frac{
 \Delta_{D} \int{\cal D}\pi {\cal D} \s {\cal D}\pi^{*} {\cal D}\s^{*}
 \prod_{n}\d (\pi_n^{*})\d (\s_n^{*})
 e^{-S[\Phi, \Phi^{*}]}}{
 \int{\cal D}\pi {\cal D} \s {\cal D}\pi^{*} {\cal D}\s^{*}
 \prod_{n}\d (\pi_n^{*})\d (\s_n^{*})
 e^{-S[\Phi, \Phi^{*}]}},
\label{4.4}
\end{eqnarray}
where the $\Delta_{D}$ is the $\Delta$-derivative for the Dirac-field
sector:
\begin{eqnarray}
\Delta_{D} &=& \sum_{n}\left(\frac{\partial^r}{\partial \V_{n}}\frac{\partial^r}{\partial \V_{n}^{*}}+\frac{\partial^r}{\partial \bar{\V}_{n}}\frac{\partial^r}{\partial \bar{\V}_{n}^{*}}
\right).
\label{4.5}
\end{eqnarray}
In order to include the contributions from the auxiliary fields in the
WT operator, we note that there is a trivial identity
\begin{eqnarray}
\int{\cal D}\pi {\cal D} \s {\cal D}\pi^{*} {\cal D}\s^{*}
 \prod_{n}\d (\pi_n^{*})\d (\s_n^{*}) 
\Delta_{X}  e^{-S[\Phi, \Phi^{*}]}=0,
\label{4.6}
\end{eqnarray}
where the $\Delta_{X}$ is the $\Delta$-derivative for the
auxiliary-field sector:
\begin{eqnarray}
 \Delta_{X} &=& -\sum_{n}\left(\frac{\partial^r}{\partial \p_{n}}\frac{\partial^r}{\partial \p_{n}^{*}}+\frac{\partial^r}{\partial \s_{n}}\frac{\partial^r}{\partial \s_{n}^{*}}
\right).
\label{4.7}
\end{eqnarray}
Let us define the WT operator for the macroscopic action
$S[\Phi,~\Phi^{*}]$ in (\ref{3.18}):
\begin{eqnarray}
\Sigma[\Phi,~\Phi^{*}] = \frac{1}{2}(S,~S) - (\Delta_{D} + \Delta_
{X})S =\frac{1}{2}(S,~S) - \Delta_{\Phi} S.
\label{4.8}
\end{eqnarray}
Adding (\ref{4.6}) to (\ref{4.4}) and using $\Delta_{\Phi}\equiv
\Delta_{D}+ \Delta_{X}$,  one finds that 
\begin{eqnarray}
\left<\Sigma [\phi, \phi^{*}]\right>_{\phi}
&=& \frac{ \int{\cal D}\pi {\cal D} \s {\cal D}\pi^{*} {\cal D}\s^{*}
 \prod_{n}\d (\pi_n^{*})\d (\s_n^{*})\Delta_{\Phi}
 e^{-S[\Phi, \Phi^{*}]}}
{ \int{\cal D}\pi {\cal D} \s {\cal D}\pi^{*} {\cal D}\s^{*}
 \prod_{n}\d (\pi_n^{*})\d (\s_n^{*})
 e^{-S[\Phi, \Phi^{*}]}}\nonumber\\
&=& \frac{ \int{\cal D}\pi {\cal D} \s {\cal D}\pi^{*} {\cal D}\s^{*}
 \prod_{n}\d (\pi_n^{*})\d (\s_n^{*})
 e^{-S[\Phi, \Phi^{*}]}\Sigma[\Phi, \Phi^{*}]}{
 \int{\cal D}\pi {\cal D} \s {\cal D}\pi^{*} {\cal D}\s^{*}
 \prod_{n}\d (\pi_n^{*})\d (\s_n^{*})
 e^{-S[\Phi, \Phi^{*}]}}\nonumber\\
&\equiv& \left<\Sigma[\Phi, \Phi^{*}]\right>_{X}.
\label{4.8} 
\end{eqnarray}
This is our fundamental relation between the WT operators in both theories.

For the microscopic theory, we have assumed that the original fermionic
action $A_{\rm c}$ is chiral invariant. This leads to the classical
master equation (\ref{3.5}). Using (\ref{3.2}) and (\ref{4.2}), one can
directly verify that $\Delta_{\phi}S_{c} =0$.  Therefore, the
microscopic action satisfies the QME,
\begin{eqnarray}
\Sigma[\phi,\phi^{*}]=0.
\label{4.9} 
\end{eqnarray}
Using (\ref{4.8}), we obtain $\left<\Sigma [\Phi,
\Phi^{*}]\right>_{X}=0$ for the macroscopic theory. This
implies the integral of the WT operator $\Sigma [\Phi, \Phi^{*}]$ over
the auxiliary fields gives zero. It is allowed a wide class of solutions
for which the WT operator becomes $\pi$ or $\s$ derivative of something. 
We consider here more restrict class of solutions for which the
macroscopic action obeys the QME 
\begin{eqnarray}
\Sigma [\Phi, \Phi^{*}]&=& \frac{1}{2}
\left(S_{D} + S_{X}, S_{D} + S_{X}\right)_{\Phi}\nonumber\\ 
&{}& - \left[\Delta_{D} + \Delta_{X}\right]\left[ S_{D} + S_{X}\right] =0. 
\label{4.10} 
\end{eqnarray}

In order to further reduce (\ref{4.10}), we assume that $\d_{B} \p_{n}$ and
 $\d_{B} \s_{n}$ are given by $C$ times functions only of $\p_{n}$ and
 $\s_n$. Since there appear no fermionic contributions in 
$\Delta_{\Phi} S$, the quantum
 master equation can be decomposed into two conditions:
\begin{eqnarray}
&{}&  \frac{1}{2}\left(S_{D}, ~S_{D}\right)_{D}
+ \left(S_{D},  S_{X}\right)_{X}=0, 
\label{4.11}\\
&{}&  \frac{1}{2} \left(S_{X},~ S_{X}\right)_{X}
- \left[\Delta_{D}S_{D}+\Delta_{X}S_{X}\right]
=0.
\label{4.12} 
\end{eqnarray}
For any macroscopic fields $\Phi^{A}$, the BRS transform $\d_{B}
\Phi^{A}$ is proportional to the ghost $C$.  It is convenient here to
introduce Grassmann even counterpart $\d$ of the odd operator $\d_{B}$.
We may define it by
\begin{eqnarray}
\d_{B} \Phi^{A} &=& - \d \Phi^{A}~ C,\nonumber\\
\Delta_{X} S_{X} &=& - \d J_{X}~ C, 
\label{4.13}
\end{eqnarray}
where $\d J_{X}$ is the change in the functional measure ${\cal 
D}\p{\cal D}\s$ induced by the chiral transformation $\d \p$ and $\d
\s$.  On the other hand, the change in the fermionic functional measure
is calculated to be
\begin{eqnarray}
\Delta_{D} S_{D} = 
2i \sum_n{\rm Tr}\left(\g_5 - \g_5 {\a}^{-1}{\tilde{D}}\right)_{nn}~ C .
\label{4.14}
\end{eqnarray}

The above relations (\ref{4.13}) and (\ref{4.14}) can be used to show
that the QME leads to the WT identities given in I.
Actually, one finds that (\ref{4.11}) and (\ref{4.12}) yield
\begin{eqnarray}
&{}& \frac{1}{2}\left(S_{D}, ~S_{D}\right)_{D}
+ \left(S_{D}, ~ S_{X}\right)_{X} \nonumber\\
&{}&~~~ = -i \bar{\V}_n\left[
 \left\{\g_5 , \tilde{D}\right\} - 2{\a}^{-1}\tilde{D}\g_5 \tilde{D} 
 - i\ \d \tilde{D}
 \right]_{nm}\V_{m}~ C =0,
\label{4.15}\\
&{}& \frac{1}{2} \left(S_{X},~ S_{X}\right)_{X} 
- \left[\Delta_{D}~ S_{D}+\Delta_{X}~ S_{X}\right]\nonumber\\
&{}&~~= - \left[\d A_{X} +2i \sum_n{\rm Tr}\left(\g_5 - \g_5 {\a}^{-1}
{\tilde{D}}\right)_{nn} - \d J_{X}\right]~C=0.
\label{4.16}
\end{eqnarray}

Having obtained the WT identities from the QME, 
we are now in a position to solve them. For notational simplicity, we
take below $\a =1$, unless otherwise stated.

\subsection{Solutions to the QME}
Let us consider first (\ref{4.15}) which reduces to   
\begin{eqnarray}
\left[\left\{\g_{5} , \tilde{D}\right\} - 2\tilde{D}\g_5 \tilde{D} 
 - i\ \d \tilde{D}
 \right]_{nm}=0 ,
\label{5.1}
\end{eqnarray}
where   
\begin{eqnarray}
\tilde{D}_{nm}&=&D_{nm}+\d_{nm}X_{n} \nn\\
X_{n}&=&(i\g_{5}\p +\s)_{n}. 
\label{5.11}
\end{eqnarray}
This is the GW relation for our system with auxiliary fields.  
As discussed in I, it is straightforward to determine $\d X_{n}$ owing
to the locality assumption: 
\begin{eqnarray}
 \left\{
 \begin{array}{ccl}
  \d X_{n}&=& -2i~\g_{5}\left(X_{n}- X_{n}X_{n}\right)\\
  \d\p_n & \equiv & 
    -2\s_n + 2\left(\s_n^2 - \p_n^2\right) \\
  \d\s_n & \equiv & 
    2\p_n - 4\s_n\p_n 
 \end{array}
 \right. .
 \label{5.2}
\end{eqnarray}
Note that $X$ commutes with $\g_{5}$, and 
obeys is a nonlinear transformation.\footnote{The matrix $\g_{5}$
satisfies $\g_{5}^{2}=1$.}
Using this result,  the GW relation reduces to 
\begin{eqnarray}
  \biggl[\left\{\g_5 ,D\right\} 
        - 2D\g_5 D - 2 D\g_5 X
        - 2 X\g_5 D
        - i\ \d D \biggr]_{nm}= 0 .
 \label{5.3} 
\end{eqnarray}

In order to solve (\ref{5.3}), we make an ansatz for the Dirac operator:
\begin{eqnarray}
 \tilde{D} 
 & \equiv &
 D_{0} + \left(1+{\cal L}(D_{0})\right) X \left(1+{\cal R}(D_{0})\right) 
   \nonumber \\
 &  = & D + X ,\nonumber  \\
 D & = & D_{0} + {\cal L}(D_{0})X +X{\cal R}(D_{0})+{\cal L}(D_{0})X{\cal R}(D_{0}),  
\label{5.4}
\end{eqnarray}
where the $ D_{0}$ is the Dirac operator in the free theory. It
satisfies the original GW relation
\begin{eqnarray}
 \{\g_5,D_{0}\} = 2 D_{0}\g_{5}D_{0}.
\label{5.5}
\end{eqnarray}
Let us suppose that a solution for $D_0$ such as the Neuberger's type
\cite{neu1} is given, and determine the functions ${\cal L} ( D_{0})$
and ${\cal R}( D_{0})$.  One substitutes (\ref{5.4}) into (\ref{5.3})
using (\ref{5.2}) for $\d X$. Then, the resulting expression for
l.h.s. of (\ref{5.3}) can be expanded in powers of $X$: There appear
linear and quadratic terms in $X$. As shown in Appendix, both of these
terms vanish if the following conditions are satisfied:
\begin{eqnarray}
 \left(1-2 D_{0}\right) \g_5  \left(1+{\cal L}\right)
  = \left(1+{\cal L}\right) \g_5,
  \label{5.60} \\
 \left(1+{\cal R}\right) \g_5\left(1-2 D_{0}\right) 
  = \g_5 \left(1+{\cal R}\right),
  \label{5.61} \\
 (1+{\cal R})\g_5 (1+{\cal L})\g_5 =\g_5 (1+{\cal L})\g_5 (1+{\cal R})=1  .
  \label{5.6}
\end{eqnarray}
We find two sets of solutions to these equations given by
\begin{eqnarray}
{\cal L}(D_0) &=& - D_0 + \ \frac{1}{1- \g_5{}D_0{}\g_5}
            \ \g_5{}D_0{}\g_5{}D_0{} \nonumber \\
        &=& -\frac{1-2 \g_5{}D_0{}\g_5}{1- \g_5{}D_0{}\g_5}
             {}D_0 , \nonumber\\
 {\cal R}(D_0) &=& - D_0 ,
\label{5.7}
\end{eqnarray}
or
\begin{eqnarray}
{\cal L}(D_0) &=& -D_0 ,
  \nonumber \\
 {\cal R}(D_0) &=& - D_0 + {}D_0{}\g_5{}D_0{}\g_5 \
            \frac{1}{1-\g_5{}D_0{}\g_5}  \label{33} \nonumber\\
        &=& - D_0 \ 
            \frac{1-2 \g_5{}D_0{}\g_5}{1- \g_5{}D_0{}\g_5}.
\label{5.8}  
\end{eqnarray}
In Appendix, we show that ${\cal L}$ and ${\cal R}$ in (\ref{5.7}) solve (\ref{5.60})
$\sim $ (\ref{5.6}).  In summary, the Dirac operator which solves the GW
relation (\ref{5.1}) is given by
\begin{eqnarray}
\tilde{D}  &=&
 D_{0} +  \g_5\left( 
            \frac{1}{1- D_0}\right)\g_5  
         \ X \ \left(1- D_0\right),
\label{5.9}
\end{eqnarray}
or
\begin{eqnarray}
\tilde{D} 
  &=& 
 D_{0} + \left(1-D_0\right)~ X~
         \g_5\left( 
            \frac{1}{1- D_0}\right)\g_5. 
\label{5.10}
\end{eqnarray}
Let us consider (\ref{5.9}) and (\ref{5.10}). Since a matrix notation is
used, the singularities arising from $D_{0}=1$ cannot be eliminated with
the factor $(1-D_{0})$. The momenta satisfying the condition $D_{0} =1$
are those at which the species doublers appear.  Therefore, the Yukawa
terms suffer from the singularities. This kind of singularities have
been discussed in ref.\cite{fi} in a different context.

Let us postpone discussion on the above singularities, and turn to the
other condition (\ref{4.16}).  It reduces to three equations:
\begin{eqnarray}
\d A^{(0)}_{X}[\p,\s]  &=& 0 ,
\label{5.11}\\
 \d A^{(1)}_{X}[\p,\s] &=& 
\d J_{X}[\p,\s] = -8\sum_{n}\p_n \nonumber \\
     &=& i 2^{2-{\frac{d}{2}}}\sum_{n}{\rm Tr}\g_5
\left(X_{n}-X^{\dagger}_{n}\right),
  \label{5.12} \\
 \d A^{(2)}_{X}[\p,\s] &=& - 2i \sum_n{\rm Tr}\left(\g_5 - \g_5 \tilde{D}
\right)_{nn}\nonumber\\
&=&
    2i \sum_n{\rm Tr}\left(\g_5(1+L)X(1+R)\right)_{nn}
\nn\\
&=&    2i \sum_n{\rm Tr}\left(\g_5 X\right)_{nn},  
\label{5.13}
\end{eqnarray} 
where we have used (\ref{5.6}) to obtain the last expression of
(\ref{5.13}).  The factor $ 2^{-d/2}$ in (\ref{5.12}) is needed to
normalize the trace, denoted by ${\rm Tr}$, over the spinor indices. In
(\ref{5.11}), $A_{X}^{(0)}$ corresponds to an invariant potential. The
terms $ A_{X}^{(1)}$ and $ A_{X}^{(2)}$ are counter terms needed to
cancel $\Delta_{X} S_{X}$ and $\Delta_{D} S_{D}$. Solutions of the above
conditions are given by
\begin{eqnarray}
A_{X}^{(0)}[\p,\s] &=& \sum_{n}h\left(\frac{\p_{n}^2 + \s_{n}^2}
     {1- 2\s_n + \p_{n}^2 + \s_{n}^2}\right) , 
  \label{5.14}\\
 A_{X}^{(1)}[\p,\s] &=& \sum_{n}2
  \ln\left(1- 2\s_n + \p_{n}^2 + \s_{n}^2\right)
\nonumber\\
  &=&   2^{1-\frac{d}{2}}\sum_{n}{\rm Tr}
   \ln\left(\left(1- X_n \right)
      \left(1- X_n^{\dag}\right)\right) , 
\label{5.15} \\
 A_{X}^{(2)}[\p,\s] & = &
 \sum_{n}{\rm Tr}\left(\ln\left(1- X\right)
                        \right)_{nn},   
\label{5.16}
\end{eqnarray}
where a function $h$ is introduced to describe the invariant potential.  
We notice that all these three terms become singular at $X=1$.

In this section, we have solved the QME (\ref{4.10})
to determine the effective action on the coarse lattice. The solutions
we have obtained are used to construct a lattice chiral symmetry. Let us 
discuss its structure in the next section.

\section{Lattice chiral symmetry in the macroscopic theory} 
Let us first summarize our results. In contrast to I, we discuss quantization
of the system in the antifield formalism.
The partition function for the macroscopic 
theory is given by
\begin{eqnarray}
 Z_{\rm MACRO}
 & = & \int {\cal D}\Phi {\cal D}\Phi^{*}\prod_{A} \d (\Phi^{*}_{A})
\exp -S[\Phi,~\Phi^{*}],
\label{6.1}
\end{eqnarray}
with the total macroscopic action  
\begin{eqnarray}
S[\Phi,~\Phi^{*}]&=&  S_{D} + S_{X},\nonumber\\
S_{X} &=&  A_{X}^{(0)} +  A_{X}^{(1)}+  A_{X}^{(2)}\nonumber\\
 &{}&+2
 \sum_{n}\left[\pi^{*}_{n}\left\{\s_n - \left(\s_n^2 - \p_n^2\right)\right\}C 
 \right.\nn \\ && \qquad \quad \left.
+ \s^{*}_{n}\left\{ -\p_n + 2\s_n\p_n\right\}C
  \right].
\label{6.2}
\end{eqnarray}
There arise four types of the action $S_{D}$ for the Dirac fields:
\begin{eqnarray}
(I)~~ S_{D} &=& \sum_{nm}\left[\bar{\V}_{n}{\tilde D}_{nm}
\V_{m} + 
\V_{n}^{*}i~C \g_{5}\left(1 -2{\tilde D}\right)_{nm}
\V_{m} -  \bar{\V}_{n} i~C \g_{5}\d_{nm} \bar{\V}_{m}^{*} 
\right],\nn\\
{\tilde D}&=&
 D_{0} +\g_5 \frac{1}{1-D_0} \g_5 \ X 
                  \left(1-D_0\right) ,
\label{SDI}
\end{eqnarray}
\begin{eqnarray}
(II)~~ S_{D} &=& \sum_{nm}\left[\bar{\V}_{n}{\tilde D}_{nm}
\V_{m} + \V_{n}^{*}i~C \g_{5}\d_{nm}
\V_{m} -  \bar{\V}_{n} i~C \left(1 -2 {\tilde D}\right)_{nm} \g_{5} 
\bar{\V}_{m}^{*} \right],\nn\\
{\tilde D}&=&
 D_{0} +\g_5 \frac{1}{1-D_0} \g_5 \ X 
                  \left(1-D_0\right) ,
\label{SDII}
\end{eqnarray}
\begin{eqnarray}
(III)~~S_{D} & = & \sum_{nm}\left[\bar{\V}_{n}{\tilde D}_{nm}
\V_{m} + \V_{n}^{*}i~C \g_{5}\left(1 -2{\tilde D}\right)_{nm}
\V_{m} -  \bar{\V}_{n} i~C \g_{5}\d_{nm} \bar{\V}_{m}^{*} 
\right],\nn\\
{\tilde D}&=& D_{0} +\left(1-D_0\right)  X 
                  \g_5 \frac{1}{1-D_0}\g_5 , 
\label{SDIII}
\end{eqnarray}
\begin{eqnarray}
(IV)~~S_{D} & = & \sum_{nm}\left[\bar{\V}_{n}{\tilde D}_{nm}
\V_{m} + \V_{n}^{*}i~C \g_{5}\d_{nm}
\V_{m} -  \bar{\V}_{n} i~C \left(1 -2 {\tilde D}\right)_{nm} \g_{5} 
\bar{\V}_{m}^{*}\right],\nn\\
{\tilde D}&=& D_{0} +\left(1-D_0\right)  X 
                  \g_5 \frac{1}{1-D_0}\g_5 . 
\label{SDIV}
\end{eqnarray}

The potential terms of the auxiliary
fields, $ A_{X}^{(0)}$, $ A_{X}^{(1)}$ and $ A_{X}^{(2)}$ are given in (\ref{5.14}) $\sim$ (\ref{5.16}). Under the BRS
transformation
\begin{eqnarray}
\d_{B} \V_{n} &=& i~C \g_{5}\left(1 -2 {\tilde D}\right)_{nm}
\V_{m} , \nonumber\\ 
\d_{B} \bar{\V}_{n} &=& i~C  \bar{\V}_{n} \g_{5},\nonumber\\ 
 \d_{B} X_{n}&=& 2i~\g_{5}\left(X_n- X_{n}X_{n}\right)~C,
\label{6.4}
\end{eqnarray}
or 
\begin{eqnarray}
\d_{B} \V_{n} &=& i~C \g_{5} \V_{n} , \nonumber\\ 
     \d_{B} \bar{\V}_{n} &=& i~C  \bar{\V}_{m} 
      \left(1 -2 {\tilde D}\right)_{mn}\g_{5},\nonumber\\ 
     \d_{B} X_{n}&=& 2i~\g_{5}\left(X_n- X_{n}X_{n}\right)~C,
\label{6.41}
\end{eqnarray}
the functional measure in (\ref{6.1}) multiplies by
the contributions from the counter action,
\begin{eqnarray}
\lefteqn{
  {\cal D}\Phi {\cal D}\Phi^{*}\prod_{A} \d (\Phi^{*}_{A})
  e^{-A_{X}^{(1)}[\p,\s] -A_{X}^{(2)}[\p,\s]}} \nonumber\\
& = & \prod_{n}d \V_{n}d \bar{\V}_{n}d \p_{n} d \s_{n}
\prod_{m} d \V_{m}^{*} d \bar{\V}_{m}^{*}d \p_{m}^{*} d \s_{m}^{*}
\d (\V_{m}^{*})\d (\bar{\V}_{m}^{*}) \d (\p_{m}^{*})\d (\s_{m}^{*})
   \nonumber \\
&{}&  ~ \times  e^{-A_{X}^{(1)}[\p,\s] -A_{X}^{(2)}[\p,\s]},
\label{6.5}
\end{eqnarray}
remains invariant. The remaining part of the action 
\begin{eqnarray}
 S_{D} + S_{X} - A_{X}^{(1)} - A_{X}^{(2)},
\nonumber
\end{eqnarray}
is also left invariant under (\ref{6.4}) or (\ref{6.41}).  In other
words, the macroscopic action (\ref{6.2}) solves the QME,
$\Sigma[\Phi,~\Phi^{*}]= (S,~S)_{\Phi}/2 - \Delta_{\Phi}S =0$, which
demonstrates the presence of an exact chiral symmetry in the quantum
system on the coarse lattice. In (\ref{6.1}), the chiral transformation
of the macroscopic fields is encoded as the BRS transformation due to
the presence of the antifields. Once the transformation rule is known,
one may eliminate the antifields by the integration to obtain the
partition function given in I.

We have shown that the QME is solved in a closed form. It should be
noted, however, that the resulting actions (\ref{SDI}) $\sim$
(\ref{SDIV}) are singular as discussed above.  Moreover, the chiral
symmetry is realized in a peculiar way.  The macroscopic fields (except
for $\bar{\V}$ or ${\V}$) transform nonlinearly: $\d_{B} \V$ or
$\d_{B}\bar{\V} $ contains $X$, and $\d_{B} X$ has a quadratic term of
$X$.  As a result, in the fermionic action $S_{D}$ given in (\ref{6.2}),
neither the kinetic term $\bar{\V}D_0\V$ nor the Yukawa coupling
$\bar{\V}(1+{\cal L})X(1+{\cal R})\V$ is chiral invariant, while their
sum $ (\bar{\V}D_0\V) + (\bar{\V}(1+{\cal L})X(1+{\cal R})\V)$ becomes
invariant. Turn to the functional measure ${\cal D}\Phi $ of the
macroscopic fields, it is not chiral invariant so that the counter terms
should be included.  Furthermore, the basic invariant made out of the
auxiliary fields is nonpolynomial as in (\ref{5.14}).

Because of these problems, we would like to reconstruct the chiral
symmetry in such a way that (1) the actions are free from the
singularities, (2) the kinetic term of the Dirac fields, the Yukawa
coupling term and the functional measure are all chiral invariant, and
(3) the auxiliary field potential becomes polynomial.  The
reconstruction of the symmetry satisfying the above conditions can be
done by employing a new set of canonical variables for each type of
actions (\ref{SDI}) $\sim$ (\ref{SDIV}). We argue that two of the
transformed actions define consistent quantum theories but the remaining
two cases may not.

\section{Reconstruction of chiral symmetry in terms of new canonical
 variables}
Let $\hat{\Phi}^{A}= \{\T,~\Tb,~{\hat X}=i\g_{5}{\hat \p} + {\hat
\s}\}$ be new fields. In the antifield formalism, we obtain the new fields by
considering a canonical transformation from $\{\Phi^{A}, \Phi_{A}^{*}\}$
to $\{{\hat \Phi}^{A}, \hat{\Phi}^{*}_{A}\}$, where 
$ \hat{\Phi}^{*}_{A}=\{\T^{*},~\Tb^{*}, \hat{X}^{*}\}$.
  The generator is given by
\begin{eqnarray}
G[\Phi, \hat{\Phi}^{*}]= \sum_{nm}\T^{*}_{n} Y(X)_{nm}\ \V_{m} 
+\sum_{n}{\bar{\V}}_{n}U(X)_{nm}\Tb_{m}^{*} +  \sum_{n}{\rm Tr}
\left[\hat{X}^{*}_{n} W(X)_{n}\right],
\label{6.6}
\end{eqnarray}
where $\hat{X}^{*}= 2^{-d/2}[-i\g_{5}\hat{\p}^{*} + \hat{\s}^{*}]$.  The
matrices $Y$, $U$ and $W$ are functions of $X$, and symmetric in the
spinor indices. The new fields are obtained by ${\hat \Phi}^{A}=\partial
G/\partial \hat{\Phi}^{*}_{A}$, while the old set of antifields are
given by $ \Phi_{A}^{*}=\partial G/\partial \Phi^{A}$ : 
\begin{eqnarray}
\T_{n} &=& \left[Y(X) \V\right]_{n} ,\nonumber\\
\Tb_{n} &=& \left[{\bar\V}U(X)\right]_{n},\nonumber\\
{\hat X}_{n} &=&  W(X)_{n}, \nonumber\\
{\V}^{*}_{n} &=& \left[\T^{*} Y(X)\right]_{n}, \nonumber\\
{\bar \V}^{*}_{n}&=& \left[U(X)\Tb^{*}\right]_{n},\nonumber\\
X^{*}_{n} &=& {\rm Tr} \left[
\hat{X}^{*}_{n}\frac{\partial W(X)_{n}}{\partial X_{n}}\right]
+ \sum_{ml}
\T^{*}_{m}\frac{\partial Y(X)_{ml}}{\partial X_{n}} \V_{l}\nn\\ 
&{}& +
\sum_{ml}{\bar \V}_{m} 
\frac{\partial Y(X)_{ml}}{\partial X_{n}}\Tb^{*}_{l} .
\label{6.7} 
\end{eqnarray}

There is a variety of choices for the matrices $Y$, $U$ and $W$ with
which the transformed actions are free from the singularities. Among
them, we discuss the following four cases corresponding to the actions
(\ref{SDI}) $\sim$ (\ref{SDIV}). 

\underline{The case (i)}: 
For (\ref{SDI}), we take
\begin{eqnarray}
Y(X)_{nm} &=& \left[\frac{1}{1 -D_{0}}(1 - X)(1 - D_{0})\right]_{nm},\nn\\
U(X)_{mn} &=& \d_{nm}, \nn\\
W(X)_{n} &=& \frac{X_{n}}{1 - X_{n}}.
\label{YUX1} 
\end{eqnarray}
One can confirm that the Jacobian factor associated with the change of
variables from $\{\pi,~\s\}_{n}$ to $\{\hat{\p}_{n},~\hat{\s}_{n}\}$
exactly cancels the contribution from the counter term
$A_{X}^{(1)}$:
\begin{eqnarray}
{\cal D} \pi {\cal D}\s~e^{-A_{X}^{(1)}} &\equiv& \prod_{n}d 
\pi_{n}d \s_{n}~e^{-A_{X}^{(1)}}\nn\\
&=& {\cal D}\hat\pi{\cal D}\hat\s.
\label{measure}
\end{eqnarray}
 Likewise, since
\begin{eqnarray}
\sum_{n}{\rm Tr} (\ln ~Y)_{nn} -A_{X}^{(2)} &=& 0,
\label{6.161}
\end{eqnarray}
one can also see that the fermionic measure with the counter term
$A_{X}^{(2)}$ becomes
\begin{eqnarray}
{\cal D} \V {\cal D}\bar{\V}~e^{-A_{X}^{(2)}} &\equiv& \prod_{n}d 
\V_{n}d \bar{\V}_{n}~e^{-A_{X}^{(2)}}
={\cal D}\T{\cal D} \Tb~{\rm Det}(Y)~e^{-A_{X}^{(2)}}\nonumber\\
&=& {\cal D}\T{\cal D} \Tb \exp  
[\sum_{n}{\rm Tr}(\ln Y)_{nn} -A_{X}^{(2)} ]\nonumber\\
&=& {\cal D}\T{\cal D} \Tb.
\label{6.15}
\end{eqnarray}
Therefore, the transformed theory is described by the partition function 
\begin{eqnarray}
Z_{\rm old }&=&
 \int {\cal D}\Phi {\cal D}\Phi^{*}\prod_{A} \d (\Phi^{*}_{A})
 \exp -(S[\Phi,~\Phi^{*}])\nonumber\\
&=& Z_{\rm new}
 = \int {\cal D}{\hat\Phi} {\cal D}{\hat\Phi}^{*}
\prod_{A} \d ({\hat\Phi}^{*}_{A})\exp -({\cal S}[{\hat\Phi},~{\hat\Phi}^{*}]), 
\label{6.18}
\end{eqnarray}
where the new action is given by
\begin{eqnarray}
{\cal S} [{\hat\Phi},~{\hat\Phi}^{*}]&=& {\cal S}_{D} + {\cal S}_{X} ,
   \nonumber\\
 {\cal S}_{D} &=& \sum_{nm} \Tb_{n} 
\left[D_{0}+ 
\hat{X}(1- D_{0})\right]_{nm}\T_{m}\nonumber\\
&{}& + 
\sum_{nm}\T^{*}_{n}C~i\g_{5} \left(1 -2 D_{0}\right)_{nm}
\T_{m}- \sum_{n}\Tb_{n}C~i\g_{5}\Tb^{*}_{n} ,
\nonumber\\ 
{\cal S}_{X} &=&  \sum_{n}h({\hat\p}_{n}^{2}+ {\hat\s}_{n}^{2})
 + 2~\sum_{n}({\hat\p}_{n}^{*}{\hat\s}_{n}-{\hat\s}_{n}^{*}{\hat\p}_{n})~C. 
\label{6.21}
\end{eqnarray}
We have used here the relations
\begin{eqnarray}
\bar{\V}_{n}\tilde{D}_{nm}\V_{m} &=& \Tb_{n} (\tilde{D} Y^{-1})_{nm}\T_{m}
 \nonumber\\
&=&  \Tb_{n} \left[D_{0} + \hat{X}(1 - D_{0})\right]_{nm}\T_{m} \nonumber\\
\frac{\p_{n}^{2} + \s_{n}^{2}}{(1-\s_{n})^{2}
+\p_{n}^{2}}&=& {\hat\p}_{n}^{2}+ {\hat\s}_{n}^{2}. 
\label{6.20}
\end{eqnarray}
The partition function (\ref{6.18}) is invariant under
\begin{eqnarray}
    \d_{B} \T_{n} &=& i~C \g_{5}\left(1 -2 D_{0}\right)_{nm}
    \T_{m} , \nonumber\\ 
    \d_{B} \bar{\T}_{n} &=& i~C  \bar{\T}_{n} \g_{5},\nonumber\\ 
    \d_{B} \hat{X}_{n}&=& 2i~\g_{5}\hat{X}_n~C.
\label{chiral1}
\end{eqnarray}
It is noticed that 
the Yukawa couplings in the action (\ref{6.21}) are the same as 
those discussed by many authors 
\cite{reviews}\cite{Naya}\cite{chand}\cite{in}\cite{fi}:   
\begin{eqnarray}
{\cal O}_{\rm Yukawa}&=&  
  \sum_{nm} \Tb_{n} 
   \left[\left(i\g_5\hat{\p}+\hat{\s}\right)(1-D_{0})\right]_{nm}
     \T_{m} \nonumber\\
&=& \sum_{nm} \bar{\V}_{n} 
   \left[\left(i\g_5{\p}+{\s}\right)(1-D_{0})\right]_{nm}
     \V_{m}\nonumber\\
&=&  
 \sum_{nm} \Tb_{n}
     i\g_5(1-D_{0})_{nm}
    \T_{m}~\hat{\p}_{n} \nn \\
&{}&
+ \sum_{nm} \Tb_{n}
     (1-D_{0})_{nm}
    \T_{m}~\hat{\s}_{n}.
\label{6.23}
\end{eqnarray}
is form invariant, and chiral invariant:
\begin{eqnarray}
\d_{B} {\cal O}_{\rm Yukawa}= ~0.
\label{6.24}
\end{eqnarray}

\underline{The case (ii)}: For (\ref{SDII}), we may choose
\begin{eqnarray}
Y(X)_{nm} &=& \d_{nm}\nn\\
U(X)_{nm} &=& \left[\g_5 \frac{1}{1-D_0}(1-X)\g_5\right]_{nm}\nn\\
W(X)_{n} &=& \frac{X_{n}}{1 - X_{n}}.
\label{YUX2} 
\end{eqnarray}
In this case, the Jacobian factor for the Dirac fields generates an
additional factor ${\rm Det}(1 - D_{0})^{-1}$, we may define the partition
function as
\begin{eqnarray}
Z_{\rm old }&=&
 \int {\cal D}\Phi {\cal D}\Phi^{*}\prod_{A} \d (\Phi^{*}_{A})
 \exp -(S[\Phi,~\Phi^{*}])\nonumber\\
&=& \frac{1}{{\rm Det} (1-D_0)}\ Z_{\rm new}\nn\\
Z_{\rm new}
 &=& \int {\cal D}{\hat\Phi} {\cal D}{\hat\Phi}^{*}
\prod_{A} \d ({\hat\Phi}^{*}_{A})\exp -({\cal S}[{\hat\Phi},~{\hat\Phi}^{*}]), 
\label{partfunc2}
\end{eqnarray}
where 
\begin{eqnarray}
{\cal S} [{\hat\Phi},~{\hat\Phi}^{*}]&=& {\cal S}_{D} + {\cal S}_{X} ,
   \nonumber\\
 {\cal S}_{D} &=& \sum_{nm} \Tb_{n} 
\left[D_{0}(1-\g_5 D_0 \g_5)\right]_{nm}\T_{m} + \sum_{nm} \Tb_{n}
\left[\hat{X}(1- \g_{5}D_{0}\g_{5}\right]_{nm}\T_{m}\nonumber\\
&{}& + 
\sum_{n}\T^{*}_{n}C~i\g_{5}\T_{m}- \sum_{n}\Tb_{n}C~i\g_{5}\Tb^{*}_{n} ,
\nonumber\\ 
{\cal S}_{X} &=&  \sum_{n}h({\hat\p}_{n}^{2}+ {\hat\s}_{n}^{2})
 + 2~\sum_{n}({\hat\p}_{n}^{*}{\hat\s}_{n}-{\hat\s}_{n}^{*}{\hat\p}_{n})~C. 
\label{action2}
\end{eqnarray}
The partition function $Z_{\rm new}$ is invariant under
\begin{eqnarray}
    \d_{B} \T_{n} &=& i~C \g_{5} \T_{n} , \nonumber\\ 
    \d_{B} \bar{\T}_{n} &=& i~C  \bar{\T}_{n} \g_{5},\nonumber\\ 
    \d_{B} \hat{X}_{n}&=& 2i~\g_{5}\hat{X}_n~C,
\label{chiral2}
\end{eqnarray}
which is the same as the standard form of the chiral transformation in
continuum (or microscopic) theories.

\underline{The case (iii)}: For the action (\ref{SDIII}), the same
results as (\ref{partfunc2}), (\ref{action2}) and (\ref{chiral2}) for
the case (ii) can be obtained with
\begin{eqnarray}
Y(X)_{nm} &=& \left[\g_5 \frac{1}{1-D_0}(1-X)\g_5\right]_{nm}\nn\\
U(X)_{nm} &=& \d_{nm} \nn\\
W(X)_{n} &=& \frac{X_{n}}{1 - X_{n}}.
\label{YUX3} 
\end{eqnarray}

\underline{The case (iv)}: For the action (\ref{SDIV}), we consider the
matrices similar to the case (i) as
\begin{eqnarray}
Y(X)_{nm} &=& \d_{nm},\nn\\
U(X)_{mn} &=& \left[(1 - D_{0})(1 - X)\frac{1}{1 -D_{0}}\right]_{nm}, \nn\\
W(X)_{n} &=& \frac{X_{n}}{1 - X_{n}}.
\label{YUX1} 
\end{eqnarray}
Then, one obtains the partition function (\ref{6.18}) with the Dirac action
\begin{eqnarray}
 {\cal S}_{D} &=& \sum_{nm} \Tb_{n} 
\left[D_{0}+ 
(1- D_{0})\hat{X}\right]_{nm}\T_{m}\nonumber\\
&{}& + 
\sum_{n}\T^{*}_{n}C~i\g_{5} 
\T_{n}- \sum_{nm}\Tb_{n}C~i\left(1 -2 D_{0}\right)_{nm}\g_{5}\Tb^{*}_{m} ,
\label{Diracaction4}
\end{eqnarray}
The chiral transformation takes of the form
\begin{eqnarray}
    \d_{B} \T_{n} &=& i~C \g_{5} \T_{n} , \nonumber\\ 
    \d_{B} \bar{\T}_{n} &=& i~C  \bar{\T}_{n} (1-2 D_0)\g_{5},\nonumber\\ 
    \d_{B} \hat{X}_{n}&=& 2i~\g_{5}\hat{X}_n~C.
\label{chiral4}
\end{eqnarray}

Let us discuss some physical consequences for the cases (i) $\sim$ (iv)
listed above. The four are classified into two two groups: {(i) and
(iv)}, {(ii) and (iii)}.  Actually, (ii) and (iii) share the same action
(\ref{action2}) and the chiral transformation (\ref{chiral2}). The
kinetic term of the Dirac fields as well as the Yukawa term in this
action contains the factor $(1- \g_{5}D_{0}\g_{5})$ in front of the
Dirac fields.  Since $D_{0}=1$ at the momenta where the doubler modes
appear, this factor vanishes. Thus, the doubler modes remain massless,
and decouple with the auxiliary fields. However, this is the case only
at tree level, and decoupling could not persist at quantum level:
There are other chiral invariant Yukawa terms such as
\begin{eqnarray}
 \bar{\C}_n \hat{X}_n \C_n , \qquad 
 \bar{\C}_n \hat{X}_n (\g_5D_{0}\g_5D_{0})_{nm} \C_m , \nn
\end{eqnarray}
in which the doublers couple with the auxiliary fields. There are no
reasons to exclude these terms in the quantum corrections. Therefore,
(ii) and (iii) cannot give a consistent theory. 

Unlike these cases, the doubler modes in (i) and (iv) are massive and
decouple with the auxiliary fields because of the factor $(1 - D_{0})$
in the Yukawa couplings. The chiral invariant Yukawa terms always
contain this factor, and therefore the chiral invariance protects the
couplings of the doublers to the auxiliary fields. This can be seen by
use of the chiral decomposition   
\cite{reviews}\cite{Naya}\cite{iis2}:
\begin{eqnarray}
\hat{\T}_R &=& \frac{1+\hat{\gamma}_5}{2} \T, \nn\\
\hat{\T}_L &=& \frac{1-\hat{\gamma}_5}{2} \T, \nn\\
\bar{\T}_R &=& \bar{\T}\frac{1+\gamma_5}{2}, \nn\\
\bar{\T}_L &=& \bar{\T}\frac{1-\gamma_5}{2},
\label{chiral-projection}
\end{eqnarray} 
where 
\begin{eqnarray}
\hat{\gamma}_{5} = \gamma_{5}(1 - 2 D_{0}).
\label{gamma5hat}
\end{eqnarray} 
Using 
\begin{eqnarray}
\delta \hat{\T}_R &=& i~C~ \hat{\T}_R,~~~~~~~
\delta \hat{\T}_L = -i~C~ \hat{\T}_L, \nn\\
\delta \bar{\T}_R &=& i~C~ \bar{\T}_R,~~~~~~~
\delta \bar{\T}_L = -i~C~\bar{\T}_L,
\label{chiral-charge}
\end{eqnarray}
we may construct the the Yukawa term by using the chiral projection. One
finds that resulting Yukawa term is exactly the same as ${\cal O}_{\rm
Yukawa}$ in (\ref{6.23}):
\begin{eqnarray}
{\cal F}_{\rm Yukawa} &\equiv& \bar{\T}_R~ \hat{\T}_R (\hat{\s} + i\hat{\pi}) +  
\bar{\T}_L~ \hat{\T}_L (\hat{\s} - i\hat{\pi}) \nn\\
&=& \bar{\T}\hat{X}(1- D_{0})\T = {\cal O}_{\rm Yukawa}
\label{Yukawa}
\end{eqnarray}

In the transformed theory, the integration over the new auxiliary fields
can be performed explicitly. The last expression of the Yukawa term in
(\ref{6.23}) can be used to do it. One then obtains
\begin{eqnarray}
 \lefteqn{ {Z_{\rm MACRO}} } \nonumber\\
 &=&
  \int{\cal D}\T{\cal D} \Tb  \ 
  \exp\left(-\sum_{nm} \Tb_{n} \left(D_{0}\right)_{nm}\T_{m}   
            -h\left({\cal O}_{\rm 4\mbox{-}fermi}
                            (\T,~ \Tb)
              \right)
      \right) ,
 \label{6.25}
\end{eqnarray}
where the antifields are integrated, too.  The four-fermi interaction
operator ${\cal O}_{\rm 4\mbox{-}fermi}$ is given by
\begin{eqnarray}
 {\cal O}_{\rm 4\mbox{-}fermi} (\T,~ \Tb)
&\equiv& 
 \left( \sum_{nm}  \Tb_{n}
      i\g_5 (1-D_{0})_{nm} \T_{m}
    \right)^2  \nonumber \\ &{}&
    +
 \left(  \sum_{nm} \Tb_{n}
    (1-D_{0})_{nm} \T_{m}
    \right)^2 . \nonumber\\
 \label{6.26}
\end{eqnarray}
For the simplest case $h(x)=x$, we obtain the Nambu-Jona-Lasinio model. 

In this section, we have reconstructed the lattice chiral symmetry using the
canonical transformations. Since these transformations are singular, the
transformed theories are only equivalent to the original ones up to the
singularities.

\section{Inclusion of the complete set of the auxiliary fields}
In the above formulation of the lattice chiral symmetry, the auxiliary
fields we have considered are restricted to a scalar and a pseudoscalar
fields. We discuss in this section inclusion of the complete set of
the auxiliary fields.  Let $\G^{i}\ (i=1\sim 2^{d})$ be the complete set 
of the Clifford algebra in $d$ (=even) dimensional space. The Dirac 
fields carry $N_{F}$ flavors, and form the fundamental representation of 
$u(N_{F})$ algebra with a basis $T^{a}\ (a=1\sim N_F^2)$ satisfying
${T^{a}}^{\dag}=T^{a}$. Let 
$\lambda^{A}\ (A=1\sim {\cal N}\equiv 2^d N_F^2)$
 be the direct product of the above two sets of the matrices,
\begin{eqnarray}
 \lambda^A \ = \ \lambda^{ia} \ = \ \G^{i} \otimes T^a ,
\label{dp}
\end{eqnarray}
normalized by
\begin{eqnarray}
 {\rm Tr}({\lambda^A}^{\dag}\lambda^B) \ =\ 
 {\rm tr}_{(\G)}{\rm tr}_{(T)}\left({\G^{i}}^{\dag}\G^{j}\right)
    \otimes \left(T^a T^b\right) 
 \ = \ 2^{\frac{d}{2}}N_F\ \d^{ij} \d^{ab} \ = \         
 2^{\frac{d}{2}}N_F\ \d^{ab}.
\label{norm}
\end{eqnarray}
Here ${\rm tr}_{(\G)}$ and ${\rm tr}_{(T)}$ denote the traces in the
spinor and the flavor spaces, respectively.

We introduce the complete
set of the auxiliary fields $x^{A}$, and define 
\begin{eqnarray}
 X &\equiv&  x^A \lambda^A ,\nn\\ 
 x^A & =& 2^{-\frac{d}{2}} N_F^{-1}
          {\rm Tr}\left({\lambda^{A}}^{\dag}X\right) ,  
\label{comX}
\end{eqnarray}
which is the extension of (\ref{5.11}).  

Let us consider the block transformation 
suppressing the antifields for simplicity. 
\begin{eqnarray}
 \lefteqn{
  \int \CD X\, 
   \exp\left({-\sum_{nm}\bar{\V}_n \tD(X)_{nm} \V_m -A_{X}[X]}\right)}
    \nn \\
 &=& 
 \int \CD \v\CD \bar{\v}\,
  \exp\left(-A_{c}[\v,\bar{\v}] 
   -\sum_{n}(\bar{\V}_n-\bar{B}_n)\a\left(\V_n-B_n\right)\right) ,
\label{funcmes}
\end{eqnarray}
where $\ds \CD X = \prod_{A=1}^{\cal N}\prod_{n} dx_{n}^{A}$. The
$\tD(X)$ is again assumed to be linear in $X$. Hereafter, we take again
$\alpha =1$. 
We may define the chiral
transformation of the macroscopic fields as
\begin{eqnarray}
 \d\V_n &=& i\g_5\left(1-2\tD\right)_{nm}\V_m ,\nn\\
 \d\bar{\V}_n &=& \bar{\V}_n i\g_5 ,\nn\\
 \d X_n &=& -i\left\{\g_5,X_n\right\} + 2i X_n\g_5 X_n .
\label{chi-X}
\end{eqnarray}
The WT identities associated with (\ref{chi-X}) are given by
\begin{eqnarray} &
 \left(i\{\g_5,\tD\}
              -2i \tD\g_5\tD +\d\tD\right)_{nm} \ =\ 0,
\label{Res-gwr1} \\ &
 \d A_{X}^{(0)}[X]\ = \ 0, 
\label{Res-gwr2} \nn \\ &  
 \d A_{X}^{(1)}[X] \ = \ \d J_{X}[X] = 
4{\cal N}\,\frac{2^{-\frac{d}{2}}}{N_F}
              \sum_{n}{\rm Tr}\left(i\g_5 X_n\right),
\label{Res-gwr3} \\ &
 \d A_{X}^{(2)}[X]  =  \d J_{\V,\bar{\V}} =
  -2i\sum_{n}{\rm Tr}\left(\g_5 -\g_5 \tD(X)\right)_{nn} ,
\label{Res-gwr4}
\end{eqnarray}
where the potential term $A_{X}$ is decomposed as
$A_{X}=A_X^{(0)}+A_X^{(1)}+A_X^{(2)}$.  It should be noted that the
chiral transformation and the WT identities essentially take the same
form as those for the truncated case, except that $X$
does not commute with $\g_{5}$ here. For the Dirac operator 
\begin{eqnarray}
 \tD(X) &=& D_0 + (1+{\cal L}(D_0))X(1+{\cal{R}}(D_0)),
\label{tD}
\end{eqnarray}
the GW relation (\ref{Res-gwr1}) can be solved with the $D_{0}$,
${\cal L}$ and ${\cal{R}}$ given in (\ref{5.6}) and (\ref{5.7}) or
(\ref{5.8}).  Likewise, one obtains the counter terms 
\begin{eqnarray}
 A_X^{(1)}[X] &=& 2{\cal N}\,\frac{2^{-\frac{d}{2}}}{N_F}
                \sum_{n}{\rm Tr}\ln\left(1-X_n \right) , \nn\\
 A_X^{(2)}[X] &=& \sum_{n}{\rm Tr}\ln\left(1-{X_n}\right)
\end{eqnarray}
 
In order to construct the invariant potential $A_{X}^{(0)}$, we may
define new set of the auxiliary fields:
\begin{eqnarray}
 \hat{X}_n &\equiv& \frac{X_n}{1- X_n } , 
\label{newX}
\end{eqnarray}
which obeys a linear transformation as
\begin{eqnarray}
 \d \hat{X}_n &=& -i\left\{\g_5,\hat{X}_n\right\}.
\label{newdX}
\end{eqnarray}
Since $\hat{X}= \hat{x}^{A} \lambda^{A}$ transforms linearly, it is easy
to obtain a quadratic invariant:
\begin{eqnarray}
 f_{\rm inv}^{I}(\hat{x}^A_n) &=& G_{AB}^I\,\hat{x}^A_n\,\hat{x}^B_n ,
 \qquad (I=1,2,\cdots ),
\label{invX}
\end{eqnarray}
where  $G_{AB}^I$ are suitable coefficients. Using these invariants, we
have
\begin{eqnarray}
 A_X^{(0)}[X] &=& \sum_{n}h\left(f_{\rm inv}^{I}(\hat{x}^A_n)\right).
\label{A0}
\end{eqnarray}      

In summary, a chiral invariant partition function $Z$ is given by
\begin{eqnarray}
 {Z} 
 &=&
 \int \left[\CD X\CD \V \CD \bar{\V}\ e^{-A_X^{(1)}-A_X^{(2)}}\right] \
  e^{-\sum_{nm}\bar{\V}_n\tD_{nm}\V_m -A_X^{(0)}},
\label{calZ}
\end{eqnarray}
where 
\begin{eqnarray}
 \tD(X) &=& D_0 + (1+{\cal L})X(1+{\cal R}) ,\nn \\
 A_X^{(0)}[X] &=& \sum_{n}h\left(f_{\rm inv}^{I}(\hat{x}^A_n)\right) ,\nn \\
 A_X^{(1)}[X]+A_X^{(2)}[X] &=& 
  \left(2{\cal N}\,\frac{2^{-\frac{d}{2}}}{N_F}+1\right)
                \sum_{n}{\rm Tr}\ln\left(1-\frac{X_n}{\a}\right) .
\label{calZ2}
\end{eqnarray}  

Let us give a special case of $d=2$ and $N_{F}=1$:
$X=i\g_5\p +\s +i\g_{\m}V_{\m}$. The chiral transformation is given by
\begin{eqnarray}
 \d\p_n &=& -2\s_n -2\p_n^2 + 2\s_n^2 + 2 {V_n^{\m}}^2,\nn\\
 \d\s_n &=& 2\p_n - 4 \p_n\s_n , \nn\\
 \d V_n^{\m} &=& -4 V^{\mu}_{n}\p_{n} .
\label{d=2}
\end{eqnarray}
The new fields $\hat{x}^{A}$ defined in (\ref{newX}) and their
transformation are given by 
\begin{eqnarray}
 \hat{\p}_n &= \ \frac{\p_n}
        {\left(1-{\s_n}\right)^2 +\left({\p_n}\right)^2
         +\left({V_n^{\m}}\right)^2},
  \quad & 
 \d\hat{\p}_n \ = \ -2\hat{\s}_n , 
         \nn\\ \nn \\
 \hat{\s}_n &= \ \frac{\s_n - \p_n^2 -\s_n^2 - {V_n^{\m}}^2}
        {\left(1-{\s_n}\right)^2 +\left({\p_n}\right)^2
         +\left({V_n^{\m}}\right)^2},
  \quad & 
 \d\hat{\s}_n \ = \ 2\hat{\p}_n , \nn\\
 \hat{V}_n^{\m} &= \ \frac{V_n^{\m}}
        {\left(1-{\s_n}\right)^2 +\left({\p_n}\right)^2
         +\left({V_n^{\m}}\right)^2},
  \quad & 
 \d\hat{V}_n^{\m} \ = 0 .
\label{d=2hatX}
\end{eqnarray}
We have two quadratic invariants:
\begin{eqnarray}
 f_{\rm inv}^1 \ = \
 \hat{\p}_n^2 + \hat{\s}_n^2 
 &=& 
 \frac{\p_n^2 +\left(\s_n -\p_n^2 -\s_n^2 - {V_n^{\m}}^2\right)^2}
      {\left[\left(1-{\s_n}\right)^2 +\left({\p_n}\right)^2
       +\left({V_n^{\m}}\right)^2\right]^2},\nn \\
 f_{\rm inv}^2 \ = \ (\hat{V}^{\m}_n)^2
 &=& 
 \frac{{V_n^{\m}}^2}
      {\left[\left(1-{\s_n}\right)^2 +\left({\p_n}\right)^2
       +\left({V_n^{\m}}\right)^2\right]^2}.
\label{d=2inv}
\end{eqnarray}

In this section, we have discussed an extension of the auxiliary method. 
We have obtained a formal expression of chiral invariant partition
function (\ref{calZ2}).  Its action, however, is not free from
the singularities discussed in section 3, and it seems to be difficult
to construct the canonical transformation which removes these
singularities.

\section{Summary and discussion}
There have been known nontrivial examples where exact chiral symmetries
are realized in interacting theories on the lattice. One was given by
L\"{u}scher who discussed chiral gauge theories.  The lattice
chiral symmetry in fermionic interacting system discussed in this paper
may provide another example.  For our fermionic system with auxiliary
fields, there arise two sets of the WT identities : One is
the GW relation which tells us how to define chiral transformation for
the Dirac as well as the auxiliary fields on the coarse lattice. Under
the suitable locality assumption for the auxiliary fields, we have
determined the chiral transformation for the macroscopic fields. The
transformation rule is used to construct chiral invariant actions. The
other WT identity can be interpreted as an anomaly matching relation
between the microscopic and the macroscopic theories. This identity
contains contributions arising from the transformation of the
functional measure, and is used to construct counter terms needed to
make the functional measure on the coarse lattice chiral invariant. In
the antifield formalism, these WT identities are obtained from the
QME $\Sigma =0$. 

Owing to the auxiliary field method, the fermionic sector of our system
is linearized. The price for it is that the integration over the
auxiliary fields remains in the condition $<\Sigma> =0$. We have considered in this
paper the QME $\Sigma =0$, and found four types of actions which solve
the QME.  However, they are found to have singularities in the Yukawa
couplings and the potential of the auxiliary fields. Those in the Yukawa
couplings are related to the presence of doubler modes. Beside these
singularities, none of the kinetic term of the Dirac fields, the Yukawa
couplings and the functional measure becomes chiral invariant in the
realization of the symmetry with the block variables.  In order to avoid
these problems, we have used more suitable sets of
variables obtained by (singular) canonical transformations. We have
discussed four types of the transformed actions. In all cases, the new
fields transform linearly, and have chiral invariant functional measure. 
Among these actions, only two of them define consistent quantum
theories. They are exactly equivalent to those obtained by using the
representation method for chiral algebra arising from free field theory:
One defines the chiral transformation in the free theory, and then
constructs chiral invariant Yukawa couplings from the chiral
decomposition. The chiral invariance in such systems is expressed by the
classical master equation rather than the QME.  Since the new auxiliary
fields belong to the conventional SO(2) multiplet, they are readily
integrated to give purely fermionic system.  The Nambu-Jona-Lasinio
model emerges as the simplest one. Our results may give justification of
the representation method for formulating lattice chiral symmetry in
theories with generic interactions.

Let us discuss some implications of our results for realization of
regularization-dependent symmetry in lattice or continuum theory. The
block transformation on the lattice or its continuum analog plays an
important r\^{o}le in inheriting symmetry properties of the macroscopic
theory from those of the microscopic theory.  In general, the functional
measure for the original block variables may not be invariant under the
cutoff-dependent symmetry transformations. The induced change in the
functional measure corresponds to the $\Delta$ derivative of the
extended action, whose explicit expression depends on the UV
regularization scheme. However, unless nontrivial anomaly is present in
the given theory, one can always find the counter action needed to
cancel the $\Delta$ derivative contribution. The counter action is
regularization dependent, but is expected to be antifield
independent. Then, a canonical transformation will be performed in such
a way that new fields have invariant measure. This corresponds to the
reduction of the QME to the classical master equation.

\section{Acknowledgments} 

This work is supported in part by the Grants-in-Aid for Scientific
Research No. 12640258, 12640259 and 13135209 from the Japan Society for
the Promotion of Science. We also thank Yukawa Institute for Theoretical 
Physics in Kyoto University for hospitality and providing its computer
fascilities.

\appendix
\section{Derivation of some formulae}
Let us first derive (\ref{5.60}), (\ref{5.61}) and (\ref{5.6}). The
reduced GW relation (\ref{5.3}) is divided into the local terms and the
nonlocal terms which depend on ${\cal L}$ and/or ${\cal R}$. The nonlocal terms becomes
\begin{eqnarray}
\begin{array}{lll}
  &{}&
  \{\g_5 ,{\cal L}X+X{\cal R}+{\cal L}X{\cal R}\}  \\ &&
  -2 D_{0}\g_5 (X+{\cal L}X+X{\cal R}+{\cal L}X{\cal R})  \\ &&
  -2 (X+{\cal L}X+X{\cal R}+{\cal L}X{\cal R})\g_5 D_{0}  \\ &&
  -2 X \g_5 ({\cal L}X+X{\cal R}+{\cal L}X{\cal R})    \\ && 
  -2 \left({\cal L}X+X{\cal R}+{\cal L}X{\cal R}\right)\g_5 X      \\ && 
  -2\left({\cal L}X+X{\cal R}+{\cal L}X{\cal R})\g_5 ({\cal L}X+X{\cal R}+{\cal L}X{\cal R}\right) \\ && 
  -2 \left(\, {\cal L}\g_5(X- X^2)+\g_5(X- X^2){\cal R} +{\cal L}\g_5(X- X^2){\cal R} \, \right) \\ &&
  = {\cal K}X + X {\cal H} + {\cal K}X{\cal R} +{\cal L}X {\cal H}-2 X{\cal G}X 
-2 X {\cal G} X{\cal R} -2 {\cal L}X {\cal G}X -2 {\cal L}X{\cal G}X{\cal R} =0.
\end{array}
\label{a1} 
\end{eqnarray}
where 
\begin{eqnarray}
{\cal K} &=& \g_5 {\cal L} - {\cal L}\g_5 -2 D_0 \g_5 -2 D_0 \g_5 {\cal L}, \nn\\
{\cal H} &=& -\g_5 {\cal R} + {\cal R}\g_5 -2 \g_5 D_0 -2 {\cal R}\g_5 D_0, \nn\\
{\cal G} &=& \g_5 {\cal L} +{\cal R}\g_5 +{\cal R}\g_5 {\cal L}.
\label{a2}
\end{eqnarray}
The conditions ${\cal K}={\cal H}={\cal G}=0$ lead to (\ref{5.60}), (\ref{5.61}) and (\ref{5.6}).

Let us show that ${\cal L}$ and ${\cal R}$ in (\ref{5.7}) are solutions of
(\ref{5.60}) $\sim$ (\ref{5.6}). ${\cal K}$ with ${\cal L}$
in (\ref{5.7}) is the GW relation is free theory (\ref{5.5}). 
${\cal H}$ with ${\cal R}$ in (\ref{5.7}) becomes
\begin{eqnarray}   
{\cal H} 
 &=& -\g_5{}D_0 -D_0\g_5 +2{}D_0{}\g_5{}D_0
    \nn \\
 & &  -{}\g_5{}D_0{}\g_5{}D_0{}\g_5 \ 
                \frac{1}{1- \g_5{}D_0{}\g_5}
              +{}D_0{}\g_5{}D_0{}\g_5 \
                \frac{1}{1- \g_5{}D_0{}\g_5}\ 
                 \g_5 \left(1-2 D_0\right) 
     \nn  \\
 &=& -{}D_0{}\g_5{}D_0 \ \frac{1}{1- \g_5{}D_0{}\g_5}
     +{}D_0{}\g_5{}D_0 \ \frac{1}{1-{}D_0} \ 
       \left(1-2 D_0\right) 
     \nn  \\
 &=& {}D_0{}\g_5{}D_0 \ \frac{1}{1- \g_5{}D_0{}\g_5}
     \left(-\left(1-{}D_0\right)
                 +\left(1-{}\g_5{}D_0{}\g_5\right)
                   \left(1-2{}D_0\right)\right)
     \frac{1}{1-{}D_0} 
    \nn \\
 &=& 0 ,
\label{a3}
\end{eqnarray}
where we have used $\g_5{}D_0{}\g_5{}D_0{}\g_5=D_0{}\g_5{}D_0$.
Finally, $\cal{G}$ turns to be 
\begin{eqnarray}   
{\cal G}&=&
 {-{}\g_5{}D_0{} + \left(-{}D_0 
            +{}D_0{}\g_5{}D_0{}\g_5 \
              \frac{1}{1- \g_5{}D_0{}\g_5}\right)\g_5}
    \nn \\
  &{}&~ +{ \left(-{}D_0 +{}D_0{}\g_5{}D_0{}\g_5 \
             \frac{1}{1- \g_5{}D_0{}\g_5}\right)
              \g_5\left(-{}D_0\right)}
    \nn  \\
 &=& {\left(-\g_5{}D_0 -D_0\g_5 +2{}D_0{}\g_5{}D_0
                  \right)} 
     \nn \\ 
 & &-{}D_0{}\g_5{}D_0   
    +{}D_0{}\g_5{}D_0{}\g_5 \ \frac{1}{1- \g_5{}D_0{}\g_5} \ 
    \g_5 
    -{}D_0{}\g_5{}D_0{}\g_5 \ \frac{1}{1- \g_5{}D_0{}\g_5} \ 
    \g_5{}D_0 
     \nn \\
 &=& {}D_0{}\g_5{}D_0{}\g_5 
      \left(-1+\frac{1}{1- \g_5{}D_0{}\g_5}\right)\g_5
     -{}D_0{}\g_5{}D_0{}\g_5 \ \frac{1}{1- \g_5{}D_0{}\g_5} \ 
       \g_5{}D_0 
     \nn \\
 &=& {}D_0{}\g_5{}D_0{}\g_5 \ \frac{1}{1- \g_5{}D_0{}\g_5} \ 
      \g_5{}D_0
     -{}D_0{}\g_5{}D_0{}\g_5 \ \frac{1}{1- \g_5{}D_0{}\g_5} \ 
       \g_5{}D_0 
     \nn \\
 &=& 0 .
\label{a4}
\end{eqnarray}


\end{document}